\documentclass[usenatbib,epsfig]{mn2e}
\usepackage{graphicx}
\usepackage{deluxetable}

\title[A contracting circumbinary ring around Ori 139-409 ]
      {A contracting circumbinary molecular ring with an inner cavity of about 140 AU 
       around Ori 139-409}
\author[Zapata et al.]{Luis A. Zapata$^1$, Peter Schilke$^{1,2}$,  
                       Paul T.P. Ho$^{3,4}$ \\
                      $^1$Max-Planck-Institut f\"{u}r Radioastronomie, Auf dem H\"ugel 69, 53121, Bonn, Germany\\
                      $^2$Physikalisches Institut, Universit\"{a}t zu K\"{o}ln, Z\"{u}lpicher Str. 77, 50937 K\"{o}ln, Germany\\
                      $^3$Harvard-Smithsonian Center for Astrophysics, 60 Garden Street, Cambridge, MA 02138, USA\\
                      $^4$Academia Sinica Institute of Astronomy and Astrophysics, Taipei, Taiwan\\}

\begin{document}

\date{Accepted --. Received --; in original form --}

\pagerange{\pageref{firstpage}--\pageref{lastpage}} \pubyear{2009}

\maketitle

\label{firstpage}

\begin{abstract}
 Sensitive and subarcsecond resolution ($\sim$ 0.7\arcsec) 
CH$_3$OH(7$_{-2,6}$ $\rightarrow$ 6$_{-2,5}$) line and 890 $\mu$m continuum
observations made with the Submillimeter Array (SMA) towards the
hot molecular circumbinary ring associated with the young multiple star 
{\it Ori 139-409} are presented.  The CH$_3$OH(7$_{-2,6}$ - 6$_{-2,5}$) 
emission from the ring is well
resolved at this angular resolution revealing an inner cavity with 
a size of about 140 AU. A LTE model of a
Keplerian disk with an inner cavity of the same size confirms 
the presence of this cavity.
Additionally, this model suggests that the circumbinary ring is contracting with 
a velocity of V$_{inf}$ $\sim$ 1.5 km s$^{-1}$
toward the binary central compact circumstellar disks reported at a wavelength of 7 mm.
 The inner central cavity seems to be formed by 
the tidal effects of the young stars in the middle of the ring.
The ring appears to be not a stationary object. Furthermore,
the infall velocity we determine is about a factor of 3 slower than the
free-fall velocity corresponding to the dynamical mass.
This would correspond to a mass accretion rate of about 10$^{-5}$ M$_\odot$/yr.
 We found that the dust emission associated with Ori 139-409 appears to be arising 
from the circumstellar disks with no strong contribution from the molecular gas ring. 
Furthermore, a simple comparison with other classical molecular
dusty rings ({\it e.g.} GG Tau, UZ Tau, and UY Aur) suggests that {\it Ori 139-409} 
could be one of the youngest circumbinary 
rings reported up to date.  
Finally, our results confirm that the circumbinary rings are actively funneling fresh 
gas material to the central compact binary 
circumstellar disks, {\it i.e.} to the protostars in the very early
phases of their evolution.
\end{abstract}


\begin{figure*}
\includegraphics[scale=0.63, angle=0]{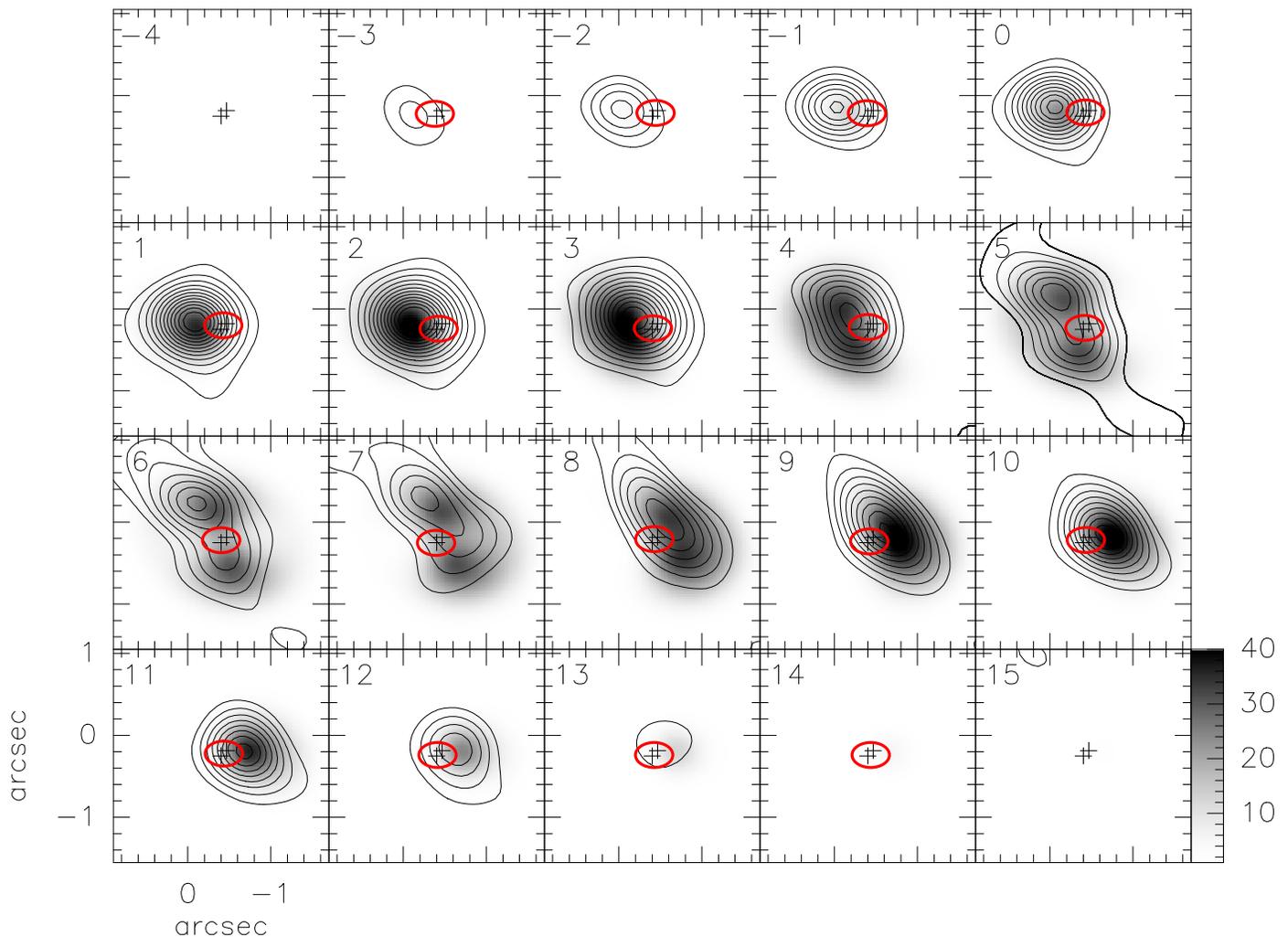}
\caption{\scriptsize Channel-velocity image of the CH$_3$OH(7$_{-2,6}$-6$_{-2,5}$) 
thermal emission from the molecular circumbinary ring {\it Ori 139-409} (contours) 
overlaid with our LTE model (grey scale). The emission is averaged
in velocity bins of 1 km s$^{-1}$. The central velocity 
is indicated (in km s$^{-1}$) on the top left-hand corner of each panel. The systemic
velocity of the ambient molecular cloud is between 6--7 km s$^{-1}$. 
The contours are 8\% to 90\% with a step of 7\% the flux peak.
The right scale grey bar indicates the T$_b$ in K.
The northern and southern extended emission located in middle channels is probably
associated with the ambient gas.
The crosses mark the position of the compact circumstellar disks \citep{Zapataetal2007}.
The red ellipse marks the position of the putative 140 AU inner cavity.}
\label{fig1}
\end{figure*}

\section{Introduction}

Most of the main sequence low-mass stars in our Galaxy 
are binaries or multiple systems, 
however its formation processes are still not
well understood. Our poor understanding about the formation of the multiple
stars has been mainly attributed to the modest sample of
relatively well studied protobinary or multiple systems at millimeter and infrared
wavelengths \citep{mo2007,Lau2004}.  
Surveys have been made at millimeter wavelengths
in an attempt to find new very young multiple systems, however these 
have been strongly limited by sensitivity \citep{dutrey2001}. Now,
Submillimeter interferometers begin to open a new ``{\it radio window}'', 
providing highly-excited molecular and
continuum observations with a better signal-to-noise ratio 
(in comparison to the millimeter observations) 
and a sub-arcsecond resolution which may be able to trace the innermost parts 
of close-by molecular and dusty circumbinary or circum-multiple disks.

{\it Ori 139-409} is part of a multiple system of compact 
millimeter and centimeter continuum sources located
in southern most part of Orion South \citep{Zapataetal2004a,Zapataetal2005,Zapataetal2007,
eis2006,eis2008,Zap2004b} and at a distance of 414 pc \citep{Mentenetal2007}.
This source is associated with a hot (T $\geq$ 100 K) and dense ($\rho$ $\simeq$ 10$^8$ cm$^3$) 
molecular circumbinary ring surrounding two
compact dusty disks (with truncated sizes of less than 50 AU) which are associated with 
intermediate mass protostars \citep{Zapataetal2007}. 
The circumbinary molecular gas ring has a deconvolved size of
$0\hbox{$.\!\!^{\prime\prime}$}64 \pm 0\hbox{$.\!\!^{\prime\prime}$}03 
\times 0\hbox{$.\!\!^{\prime\prime}$} 45 \pm 0\hbox{$.\!\!^{\prime\prime}$}04$ 
(or $\rm 294~AU \pm 14~AU \times 207~AU \pm 18$ AU) with a PA of 87$^\circ\pm 7^\circ$. 

\citet{zapa2009} reported a highly collimated SO(6$_5$$\rightarrow$5$_4$) molecular 
 and monopolar outflow with a northeast-southwest orientation 
ejected from {\it Ori 139-409} and that forms part of a more extended (2$'$) monopolar 
CO(2$\rightarrow$1) outflow that emanates from this region
\citep{Schmid-Burgketal1990}.   

In this paper, we present submillimeter and subarcsecond line and continuum 
observations made with the Submillimeter Array toward the protobinary system {\it Ori 139-409}.
We find that the molecular emission is well resolved at this angular 
resolution, revealing an inner cavity at the center of the molecular structure
with a size of about 140 AU. A LTE model of a Keplerian disk with a central hole 
agrees well with the line observations and further reveals that the circumbinary 
ring is infalling towards the compact circumstellar disks. 

\section{Observations}

Observations were made with the Submillimeter Array (SMA)\footnote{The
Submillimeter Array is a joint project between the Smithsonian
Astrophysical Observatory and the Academia Sinica Institute of
Astronomy and Astrophysics, and is funded by the Smithsonian
Institution and the Academia Sinica.} during 2008 January 14. The SMA
was in its extended configuration, which included 21 independent
baselines ranging in projected length from 30 to 258 m. The phase
reference center of the observations was R.A. = 05$^h$35$^m$13.60$^s$,
decl.= -05$^\circ$24$'$11.0$''$ (J2000.0).  The size of the primary
beam response at this frequency is about 30$''$.  The receivers were
tuned to a frequency of 349.125135 GHz in the upper sideband (USB),
while the lower sideband (LSB) was centered at 339.125135 GHz.

\begin{figure*}
\begin{center}
\includegraphics[scale=0.32, angle=0]{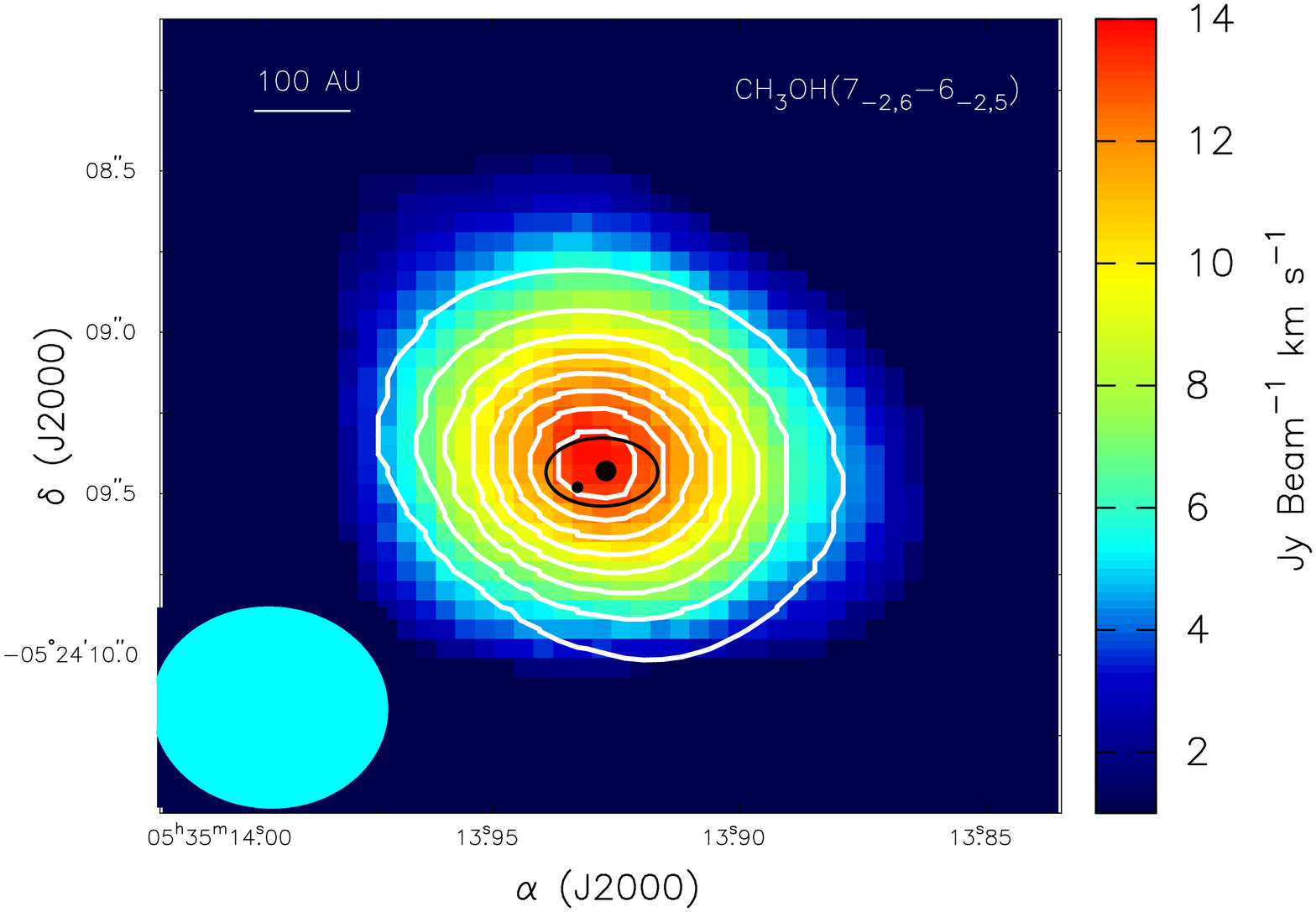}
\includegraphics[scale=0.32, angle=0]{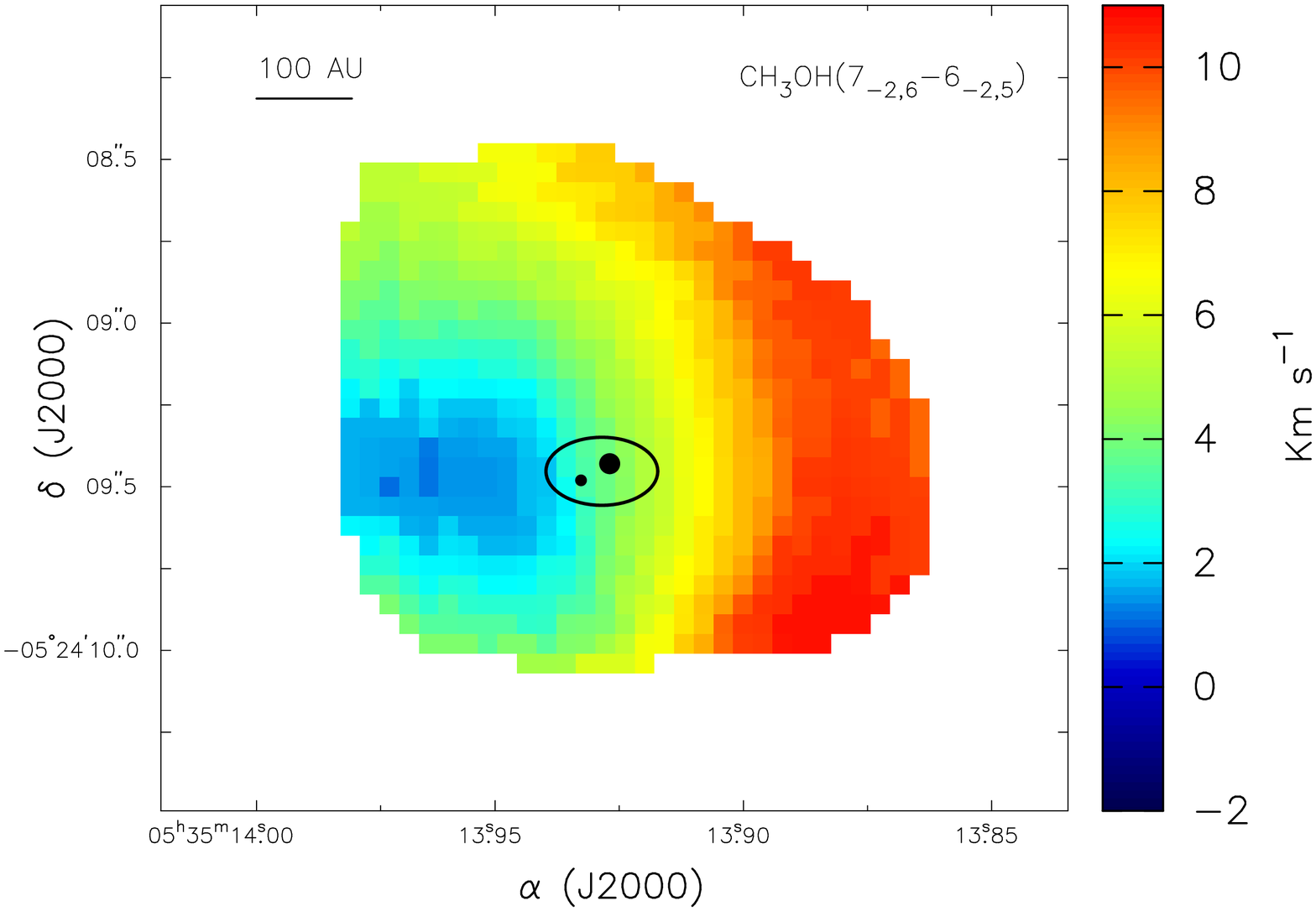}\\
\includegraphics[scale=0.32, angle=0]{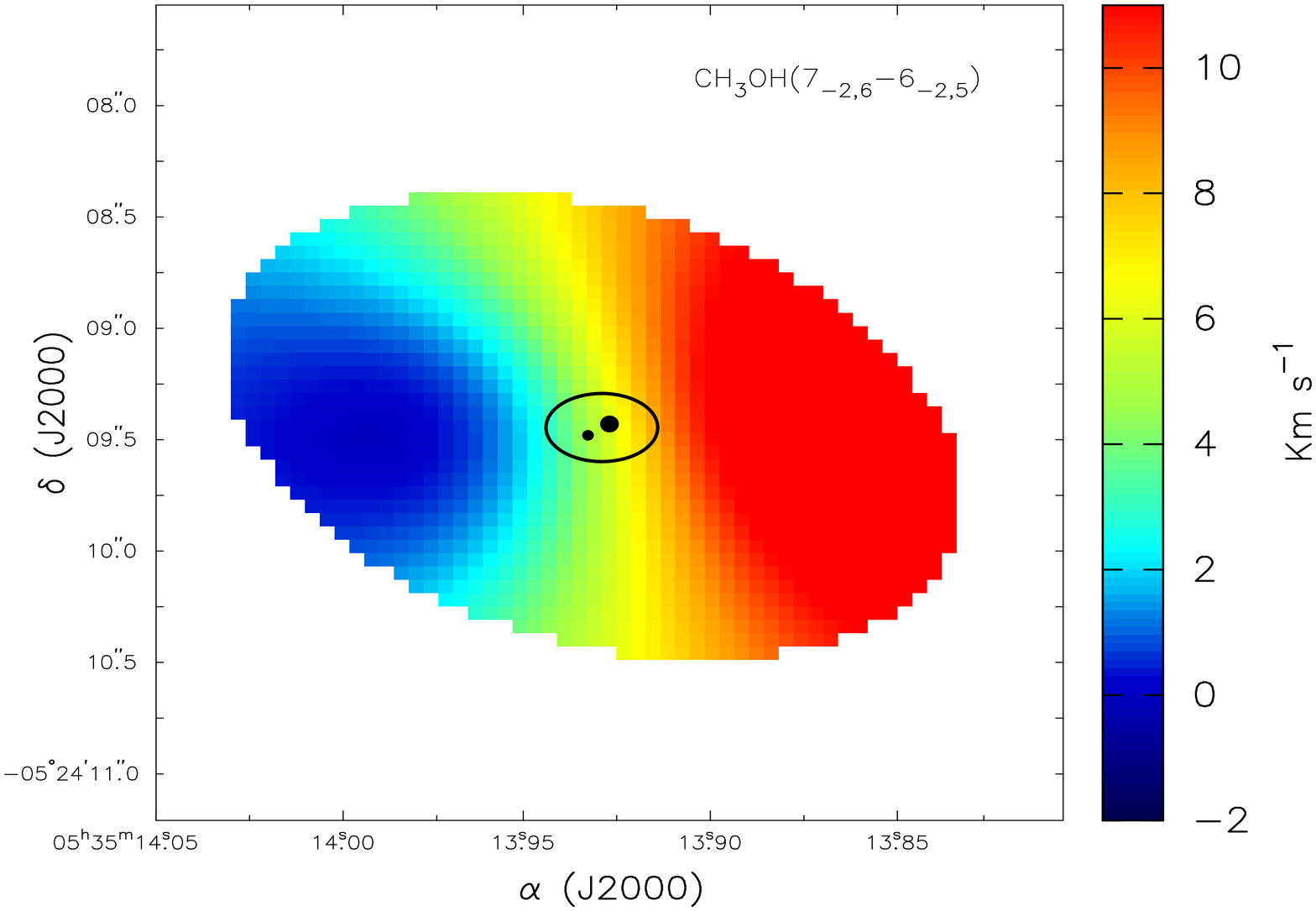}\\
\caption{\scriptsize Top Left: SMA CH$_3$OH(7$_{-2,6}$-6$_{-2,5}$) moment zero (color) 
and 890 $\mu$m continuum image (white contours) 
of the circumbinary ring. The scale bar indicates the line
integrated emission in Jy beam$^-1$ km s$^{-1}$.  The
contours are -5, 5, 10, 15, 20, 25, 30, 35, and 40
times 0.015 Jy Beam$^{-1}$, the rms noise of the image. The
synthesized beam is shown in the bottom left-hand corner
 and has a size of 0.72$''$ $\times$ 0.62$''$ with a P.A. of
87.3$^\circ$. Top Right: SMA CH$_3$OH(7$_{-2,6}$-6$_{-2,5}$) moment one
image of the circumbinary ring. The scale bar indicates the LSR radial
velocities of the disk in km s$^{-1}$. The velocity resolution was
smoothed to 1 km s$^{-1}$.  The approximate sizes and positions of the
binary circumstellar compact disks reported by \citet{Zapataetal2007} and the
presumed inner cavitiy are marked in all figures with black dots and
an ellipse, respectively. Bottom: CH$_3$OH(7$_{-2,6}$-6$_{-2,5}$) moment one
image from our LTE model. The moment one from our observations and the model
show a reasonable correspondence.}
\end{center}
\label{fig2}
\end{figure*}

The CH$_3$OH(7$_{-2,6}$ -  6$_{-2,5}$) molecule was detected in the LSB
at a frequency of 338.72 GHz. The full bandwidth of the SMA
correlator is 4 GHz (2 GHz in each band).  The SMA digital correlator
was configured in 24 spectral windows (``chunks'') of 104 MHz each,
with 256 channels distributed over each spectral window, providing a
spectral resolution of 0.40 MHz (0.35 km s$^{-1}$) per channel. However, in
this work we smoothed our spectral resolution to 1 km s$^{-1}$ per channel.

The zenith opacity ($\tau_{230 GHz}$), measured with the NRAO tipping
radiometer located at the Caltech Submillimeter Observatory, was
$\sim$ 0.06, indicating excellent weather conditions during the
observations.  Observations of MWC349 (with an adopted flux density at
this wavelength of 2.4 Jy) provided the absolute scale for the flux
density calibration.  Phase and amplitude calibrators were the quasars
0530+135 and 0541-056, with measured flux densities of 1.5 $\pm$ 0.1
and 0.7 $\pm$ 0.1 Jy, respectively.  Further technical descriptions
of the SMA and its calibration schemes can be found in
\citet{Hoetal2004}.

The data were calibrated using the IDL superset MIR, originally
developed for the Owens Valley Radio Observatory
\citep{Scovilleetal1993} and adapted for the SMA.\footnote{The MIR
cookbook by C.  Qi can be found at
http://cfa-www.harvard.edu/$\sim$cqi/mircook.html} The calibrated data
were imaged and analyzed in a standard manner using the KARMA, MIRIAD, and AIPS
packages.  We used the ROBUST parameter set to 0 to obtain an optimal compromise
between sensitivity  and angular resolution.  The continuum and line images rms
noises were 15 and 60 mJy beam$^{-1}$, respectively, at an angular resolution
of $0\rlap.{''}72$ $\times$ $0\rlap.{''}62$ with a P.A. =87.3$^\circ$.

\begin{table*}
 \centering
 \begin{minipage}{140mm}
  \caption{Observational Parameters of the Circumbinary Ring}
  \begin{tabular}{lccccc}
  \hline\hline
 Parameter & Peak Flux &  Flux Density & Deconv. Size & Deconv. P.A.\\
           & [mJy]$^a$      & [mJy Beam$^{-1}$]\footnote{For the line emission the 
           units are mJy Beam$^{-1}$ km s$^{-1}$} & [arcsec, arcsec] & [Degrees]\\
  \hline
1.3 cm       &      --         &  1.03$\pm$0.06\footnote{Values obtained from Zapata et al. (2004b)}  
                                                                       &       --        &   -- \\        
7 mm         &   3.0$\pm$1     &   6.0$\pm$1     &  $\leq$ 0.1 $\times$ $\leq$ 0.1 & -- \\
1300 $\mu$m  &   150$\pm$50    &   350$\pm$50    &  1.4$\times$0.9 & -- \\   
890  $\mu$m  &   600$\pm$100   &  800$\pm$100    &  0.4$\times$0.2 & 43 $\pm$ 22\\ 
CH$_3$OH(7$_{-2,6}$-6$_{-2,5}$) & 13.9 $\pm$0.9 $\times$ 10$^3$ & 37.0$\pm$0.9 $\times$ 10$^3$ 
& 0.9$\times$0.8 & 60 $\pm$ 44\\ 
CH$_3$CN(12$_4$-11$_4$)\footnote{Values obtained from Zapata et al. (2007)} 
& 1.5$\pm$0.1 $\times$ 10$^3$  & 1.8$\pm$0.1 $\times$ 10$^3$ & 0.6$\times$0.4 & 87 $\pm$ 7 \\
\hline\hline
\end{tabular}
\end{minipage}
\end{table*}

\section{Results}

Figure \ref{fig1} shows a channel-velocity map of the
CH$_3$OH(7$_{-2,6}$ -  6$_{-2,5}$)   
line emission from the circumbinary ring {\it Ori139-409} overlaid with 
our LTE model of an
infalling Keplerian disk with a central cavity. 
The radial velocities in the channel map
are from $-$4 to $+$15 km s$^{-1}$, with the systemic
velocity of the ambient molecular cloud of about $+$6 km s$^{-1}$. 
The CH$_3$OH(7$_{-2,6}$ -  6$_{-2,5}$) line emission from the circumbinary ring starts in the east
at a radial velocity of $-$3 km s$^{-1}$, and moves to the west, finishes at a radial 
velocity of $+$13 km s$^{-1}$. 
The redshifted molecular gas emission is found towards the west, while blueshifted one is in the east. 
The total linewidth is about $+$16 km. The radial velocity values are in very good agreement
with those reported by \citet{Zapataetal2007}. 
Approximately in the ambient velocity channels one can see the puative   
central cavity starting at a radial velocity of about $+$3 km s$^{-1}$ and finishing  
at a velocity around of $+$9 km s$^{-1}$. The central cavity is located around the position 
of compact circumstellar dusty disks reported by \citet{Zapataetal2007} at a wavelength of 7 mm.
As the circumbinary ring has an inclination $\sim$ 40$^\circ$ and a P.A. of 87$^\circ$, 
the suspected cavity shows to be very enlogated from the east to west (Figure 1 and 2)
and with limb-brightening at the edges.

The free-line continuum emission at 890 $\mu$m, the moment zero (integrated intensity), and
one (integrated velocity) of the CH$_3$OH(7$_{-2,6}$ -  6$_{-2,5}$) 
line emission from the circumbinary ring {\it Ori 139-409} are presented in Figure 2.  
The continuum and the molecular emission are well coincident and show to be 
very compact and centered in the position of the circumstellar disks.  
However, the deconvolved size for the 890 $\mu$m continuum emission shows to be more compact 
compared to that the one of the molecular emission (see Table 1). This suggests that probably the continuum emission 
could be arising only from the central circumstellar disks rather from circumbinary gas ring. We note that the small
inner cavity of the molecular gas ring is not clearly observed in the moment zero map,
only in the channel maps. 

The deconvolved sizes for the line observations at this frequency are given in Table 1.
The size for the CH$_3$OH(7$_{-2,6}$ -  6$_{-2,5}$) line emission agrees 
well with that one reported by \citet{Zapataetal2007} for the line 
CH$_3$CN(12$_4$ - 11$_4$), see also Table 1. Note that the estimation of the P.A. in the CH3OH integrated 
emission is not well determined.
The moment one map of the CH$_3$OH(7$_{-2,6}$-6$_{-2,5}$)
emission shows approximately the same east-west velocity gradient 
centered around V$_{LSR}$= $+$6 km s$^{-1}$ and reported  
by observations of the CH$_3$CN(12$_4$ -  11$_4$).
This gradient does not show a classical
butterfly morphology produced by the Keplerian velocity fields in a flattened disk 
\citep[see for some examples of classical Keplerian disks: ][]{Simon2000}.
This might be possibly due to the circumbinary molecular ring is infalling  
towards the central circumstellar disks and the resulting 
velocity fields in circumbinary ring are a combination of both, 
the infall and the Keplerian velocities. This is also observed in our LTE model,
see Figure 2. 

It is interesting to mention that similar molecular structures
to these traced by the CH$_3$OH at different radial velocities (Figure 1) were found in other high density tracers 
(SO$_2$ and CH$_3$CN) argumenting in favor of the presence of a contracting Keplerian ring
in Ori 139-409 with an inner cavity.

\begin{figure}
\begin{center}
\includegraphics[scale=0.4, angle=0]{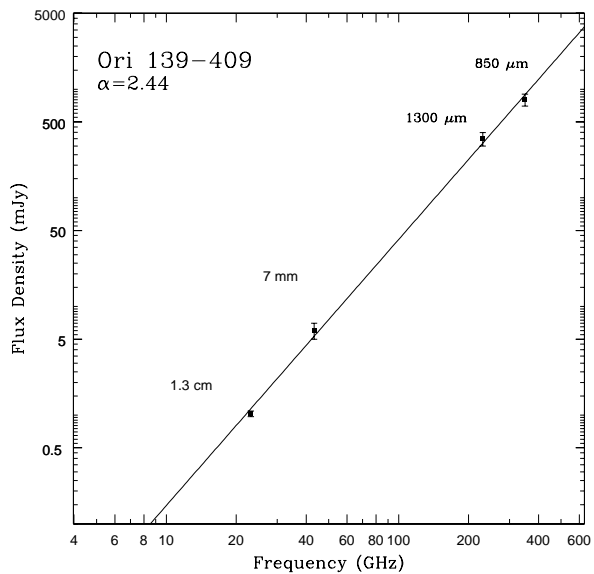}
\caption{\scriptsize SED for the source Ori 139-409 combining 1.3
cm, 7 mm, 1300 $\mu$m, and 850 $\mu$m continuum data.  
The line is a least-squares power-law fit (of the form S$_\nu \propto
\nu^\alpha$) to the spectrum. The data is well fitted by a single power-law
function.}
\label{fig3}
\end{center}
\end{figure}

We show the kinematics of the CH$_3$OH(7$_{-2,6}$ - 6$_{-2,5}$) 
gas from the circumbinary molecular ring in Figure 3. 
We have made two positional-velocity diagrams, one at a P.A. of 90$^\circ$ and 
the other one at 0$^\circ$.  The position-velocity diagrams additionally are overlaid 
with the pv-diagrams obtained from our LTE model at same angle. Both pv-diagrams, the observed and the modeled 
show a good agreement in all cases. The central putative cavity is also observed in both diagrams. 

The spectral energy distribution (SED) of {\it Ori 139-409}
from cm to sub-mm wavelengths is presented in Figure 3. In this figure one can see that 
the data is well fitted by a single power-law with $\alpha$ = 2.44 or $\beta$=0.44 (suggesting grain growth).
The value of $\alpha$ is in good agreement with that one found by \citet{Zapataetal2007}
estimated only using 7 mm and 1300 $\mu$m measurements. Note that angular resolution for the four
millimeter and centimeter observations have similar values, 1.3 cm $\sim$ 0.3$''$, 7 mm $\sim$ 0.5$''$ 
\citep[we tappered the VLA data presented by][to this angular resolution]{Zapataetal2007}, 
1300 $\mu$m $\sim$ 1$''$, and at 850 $\mu$m $\sim$ 0.7$''$.

\begin{figure}
\begin{center}
\includegraphics[scale=0.32, angle=-90]{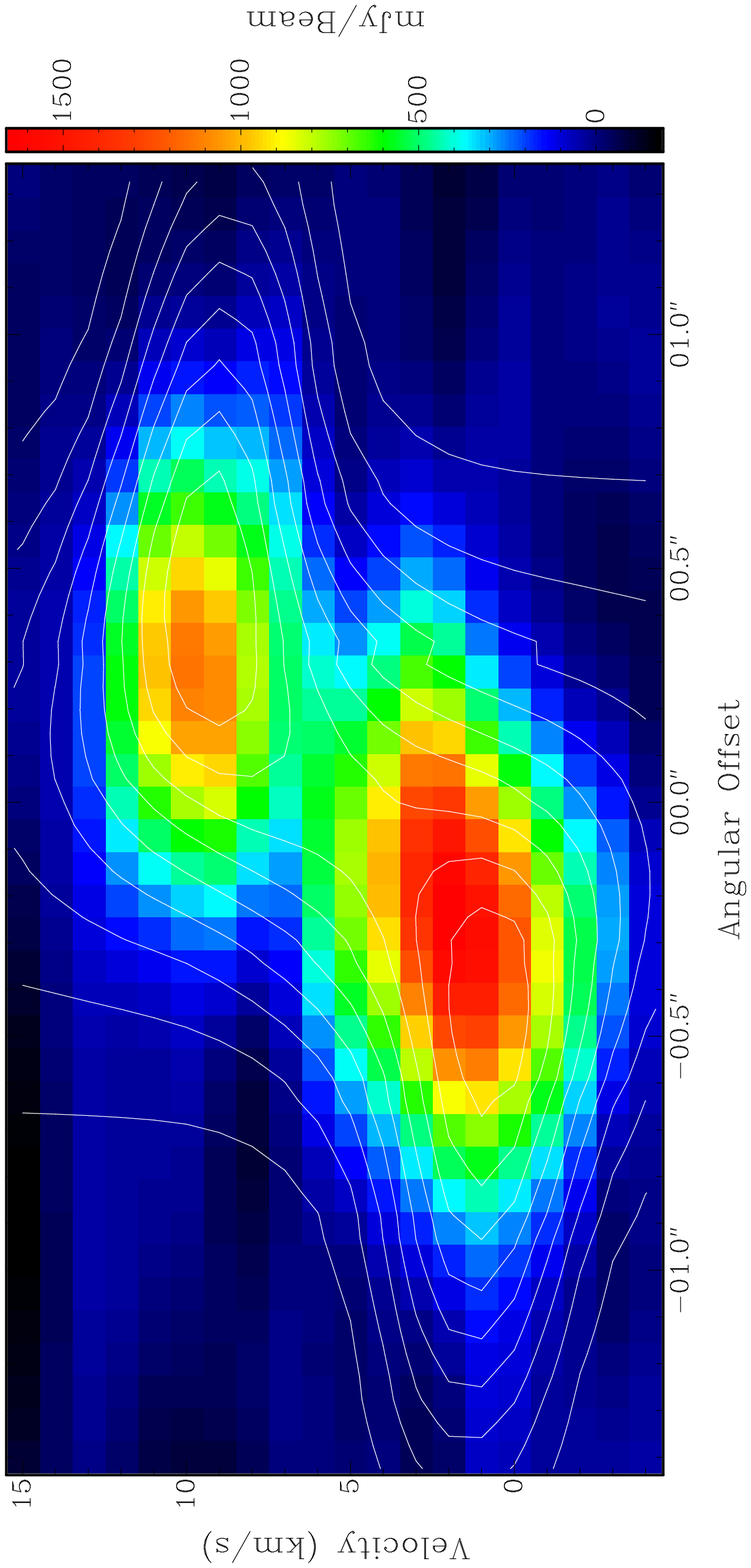}\\
\includegraphics[scale=0.31, angle=-90]{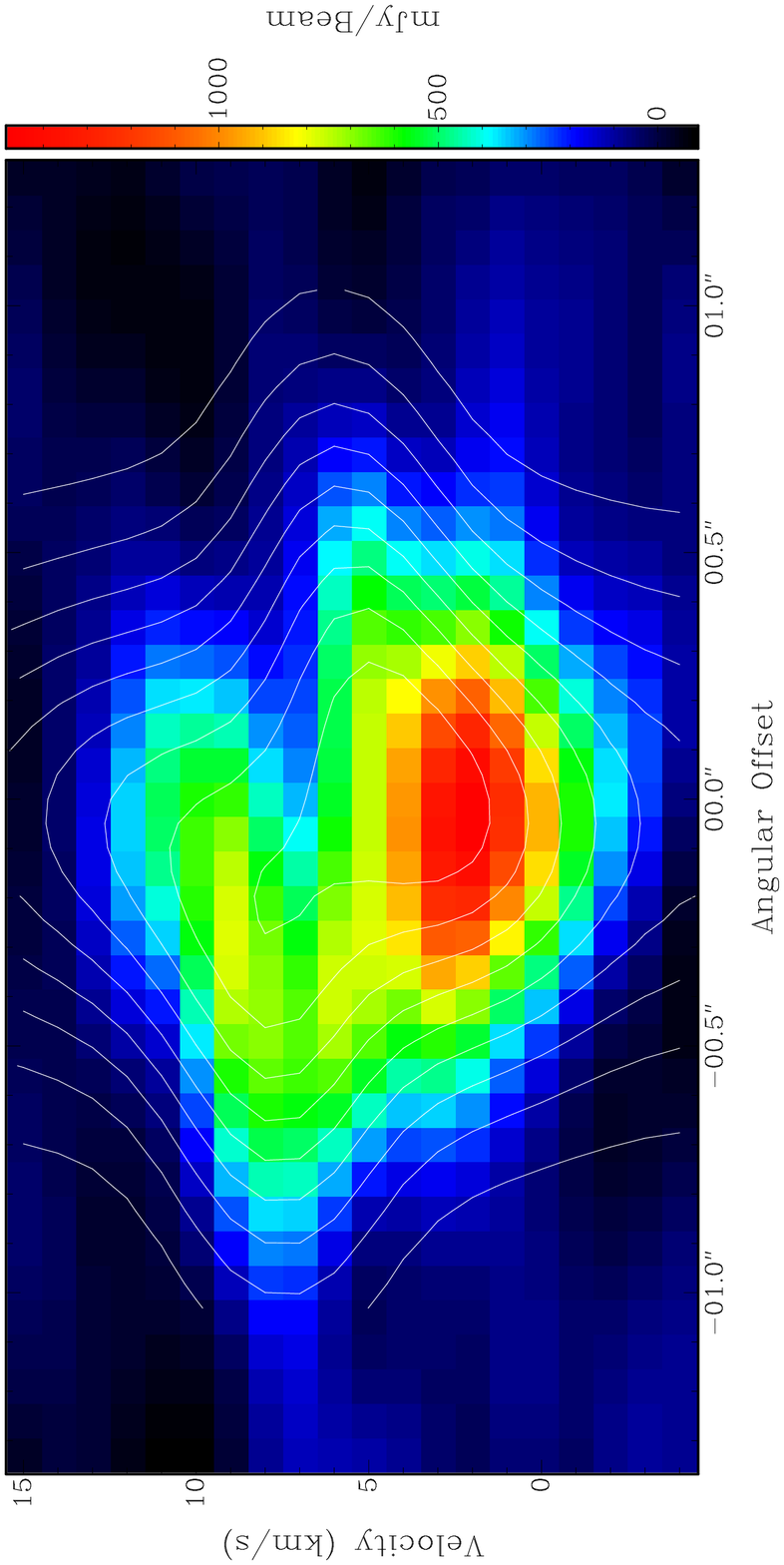}
\caption{\scriptsize Position-velocity diagrams of the CH$_3$OH(7$_{-2,6}$-6$_{-
2,5}$) line emission from the circumbinary ring,
computed with at a positional angle of 90$^\circ$ (upper panel) and 0$^\circ$ (lower panel).
In the both panels, the pv-diagrams (color scale) are compared with the pv from 
our LTE model at the same PAs (contours).
The scale bar indicates the line emission in mJy beam$^{-1}$. In both
panels, the contours are from 10\% to 90\% in steps of 10\% the flux peak
of our model. The velocity and angular resolutions are 1 km s$^{-1}$ and $\sim$
0.7$''$, respectively. The systemic velocity of the ambient molecular
cloud is 6--7 km s$^{-1}$. The horizontal axis has units of arcsecs.
The north in the lower panel is toward the left of the image, while that in the upper
panel the east is in the left.}
\label{fig4}
\end{center}
\end{figure}

\section{The Model}

We have modeled a contracting flattened disk with a central hole in
Keplerian rotation, using the disk parametrization from
\citet{Gui1999}. The contraction is assumed to have the
functional form of free-fall, i.e. $v_{\rm inf} \propto 1/\sqrt{r}$,
with a reference velocity at the reference radius, but our resolution
is insufficient to determine the exact functional form. 

The model is for the same transition CH$_3$OH(7$_{-2,6}$-6$_{- 2,5}$),
using the physical parameters presented in Table 2.  The results are
shown in Figure 1, 3 and 4.  In the model we assumed that the line
emission from the disk is in local thermodynamic equilibrium (LTE),
and took from the literature and from our observations some
determined physical values ({\it e.g.} the deconvolved size, systemic
velocity, the positional angle). The better fit than all other trials 
in our recurrence was found 
until we obtained similar structures in our model to those imaged
(see, Figures 1 and 4, and Table 2).  Most of the physical parameters
in the model were constrained in the process.  The model fits the
observations reasonably well.  A model of a Keplerian disk with the absence of an
inner cavity was also considered, however, the molecular structures in
the disk observed close to the ambient velocities were not matched
well. The dynamical mass obtained from our model is in good
agreement with that one estimated in \citet{Zapataetal2007}.

The infall velocity we determine is about a factor of 3 slower than the
free-fall velocity corresponding to the dynamical mass.  This would
correspond to a mass accretion rate of about 10$^{-5}$ M$_\odot$/yr,
and the timescale for crossing the distance from the outer to the
inner radius is in the same order as the time for a revolution in
Keplerian rotation.  The disk or ring is therefore not a stationary
object. In the model, we assumed 
that southern part of the disk is closer to us, while the northern one is
farther. This is in very good agreement with the orientation of the southwest redshifted 
monopolar outflow ejected from this source (Zapata et al 2009).

\begin{table*}
 \centering
 \begin{minipage}{100mm}
  \caption{Parameters for the molecular gas Keplerian ring LTE model}
  \begin{tabular}{lcc}
  \hline\hline
 Name & Parameter & Value\footnote{The errors in most of the physical parameters were obtained varying our model. }\\
  \hline
Systemic Velocity &  V$_{LSR} $ &  6.0 -- 7.0 km s$^{-1}$\\
Orientation & PA & 120$^\circ$ $\pm$ 5$^\circ$\\
Inclination & {\it i}&  40$^\circ$ $\pm$ 10$^\circ$\\
Density at reference radius& $n(\rm{H}_2)_0$  & 10$^8$ cm$^{-3}$\\
Size of the cavity & R$_i$ & 140 $\pm$ 30 AU\\
Dust Temperature & T$_d$ & 100 K\\
Dust Exponent & $\beta$ & 0.44\\
Distance & D & 414 pc\footnote{\citet{Mentenetal2007}}\\
Reference radius & r & 720 AU \\
Power law index density & $\omega$ & 2.75 \\
Kinetic temperature & & \\
at reference radius & T$_{kin}$  & 100 K \\
Power law index   & $\gamma$ & 0.6 \\
Scale Height of disk & H & 20 AU\footnote{Here, we assumed H constant.}\\
Dynamical mass & M & 9 M$_\odot$ $\pm$ 2 M$_\odot$\\
Infall velocity & V$_{inf}$ & 1.5 km s$^{-1}$ $\pm$ 0.5 km s$^{-1}$\\
\hline
\hline
\end{tabular}
\end{minipage}
\end{table*}

 \section{Discussion}

Our Submillimeter interferometric observations together with the LTE model revealed that the circumbinary
ring surrounding the protobinary {\it Ori 139-409} is infalling toward the central compact 
circumstellar disks.  Moreover, the channel-velocity maps of the CH$_3$OH(7$_{-2,6}$ - 6$_{-2,5}$) line emission  
show the presence of an inner cavity centered where the circumstellar dusty disks are located.
Our model of an infalling and Keplerian disk with a central cavity seems to confirm this hypothesis (see Figure 1).  
The size of this cavity is estimated to be of about 140 AU $\pm$ 20 AU. 

There are a few mechanisms that would have formed the inner cavity observed in 
the circumbinary ring. One of them would be due to the dynamic interaction of the circumbinary disk and the binary 
circumstellar compact disks located in its center. When a multiple system is formed the tidal effects of the stars
creates gaps or holes in the center of the circumbinary disks \citep{mo2007}. This is thought to be the case for the circumbinary rings:
GG Tau, SR 24 and UY Aur, see \citet[][]{Gui1999,Duv1998,And2005}. On the other hand, photo-dissociation or opacity effects
of this molecule (CH$_3$OH) close to the central binary protostars could be other possibilities.  
The photo-evaporation in flattened molecular gas disks is expected to be very important principally in the formation
of massive stars because of their strong UV fields. However, as we have a protobinary system with an enclosed dynamical 
mass of 9 M$_\odot$ (or say  about 4 M$_\odot$ for each star), this seems not to be an likely alternative. Moreover,
the SED for Ori 139-409 at centimeter wavelengths does not show any contribution of free-free emission (Figure 3).   

It is interesting to note that dust emission arising from {\it Ori 139-409} is well fitted
by a single power-law with $\alpha$ = 2.44 as obtained in our Figure 3. 
The continuum emission at 1.3 cm and 7 mm arises completely from the circumstellar 
disks as reported by \citet[][]{Zapataetal2007} and suggested by \citet[][]{Zapataetal2004a}. We additionally tapered the 7 mm 
continuum data from \citet[][]{Zapataetal2007} at an angular resolution similar to these observations 
and did not detected any emission from the ring.
If such contribution was stronger one then might be expect a power-law
with two components (one associated with the circumstellar disks obtained with a high angular resolution, and 
the other component with both contributions from the circumbinary ring and the compact disks), 
but this is not the case for {\it Ori 139-409}.    
Furthermore, the deconvolved size at 850 $\mu$m of the circumbinary 
ring is smaller as compared with the one of the CH$_3$OH(7$_{-2,6}$-6$_{-2,5}$) line
emission (see Table 1). Assuming that the emission at 850 $\mu$m is optically thin, 
isothermal (100 K) and a gas-to-dust ratio of 100, we estimated an enclosed mass of 0.03 $M_\odot$
for the source detected at this wavelength.
This value is similar to the masses of the circumstellar disks reported at 7 mm, suggesting 
that we are seeing thermal emission from the disks.  These arguments thus suggest that the continuum emission
seems to be only originate from the dusty circumstellar disks with not strong contribution from the circumbinary 
molecular gas ring. This physical phenomenon might be also the case
for the circumbinary disk in the multiple star system SR 24N, where the circumbinary disk is only detected in the CO line emission
and not in the millimeter continuum \citep[][]{And2005}. The dust emission from the circumstellar disks also appears
to be very faint. In the case of GG tau, the mm. dust emission is detected in the ring as well as in the position
of the circumstellar disks, see \citet[][]{Gui1999}.   
 
Both position-velocity diagrams made at different angles show a good agreement (Figure 4).  
In the top image of Figure 4 it is also clear that the kinematics 
of the molecular gas in the ring could be also fitted by a rigid body law, 
as suggested in Zapata et al. (2007).

If one do a simple comparison with other well known protobinary systems as GG Tau, UZ Tau, and UY Aur, one notes that the
protobinary system {\it Ori 139-409} could be one of the youngest protobinary system reported up to now. 
A clear example is that all the former multiple
systems have counterparts at infrared wavelengths associated with the central protobinary system and/or the circumbinary 
disks \citep[][]{Ro1996,Ghez1994,Hio2007} suggesting that they are likely class I or later type objects. This is contrary
to what is observed in {\it Ori 139-409} where the infrared 
emission (2.2, 8.8 and 11.7 $\mu$m ) is not detected at all \citep[][]{Gau1998,smi2004}.  
{\it Ori 139-409} is still very embedded in its parent molecular cloud and therefore is related with a true class 0 object.   
Moreover, there are a few pronounced differences between the circumbinary disk observed in GG Tau, UZ Tau, and UY Aur
and that one observed in the {\it Ori 139-409}. The millimeter continuum emission from those circumbinary disks in 
Taurus and Auriga complexes are very strong compared to that observed in {\it Ori 139-409}, 
while the molecular emission seems to go in the other sense, with the molecular emission from {\it Ori 139-409} stronger. 

The size of the cavity found in the Ori 139-409 system is in good agreement with the size of the cavity 
suggested for the L1551 binary system (of about 180 AU for $q=0.4$) estimated from theoretical models by \citep{Picha2005}. 
The L1551 binary system appears to be a very similar young stellar object to Ori 139-409. 
Both binary systems have de-projected separations of 50--80 AU, semi-major disk axes of about 10-40 AU, 
and masses of the components of approximately 0.05 M$_\odot$. A possible reason of why the circumbinary 
cavity in Ori 139-409 is something smaller than the one in L1551 could due to the former
is very young and the stars in the center have not time to made the final size 
of the cavity.         

\section{Summary}

The circumbinary ring {\it Ori 139-409}  presents a promising laboratory for future
studies formation of multiple stars. We have observed the 
($\sim$ 0.7\arcsec) CH$_3$OH(7$_{-2,6}$ - 6$_{-2,5}$) line and 890 $\mu$m continuum
emission from this young binary star with the Submillimeter Array. 
Additionally, we have modeled the circumbinary
disk associated with {\it Ori 139-409} as a molecular Kleplerian and infalling flattened disk 
with a central cavity. We have found the following results:   

\begin{itemize}

\item  The CH$_3$OH(7$_{-2,6}$ - 6$_{-2,5}$) 
emission from the ring is well
resolved at this angular resolution revealing an inner cavity with 
a size of about 140 AU; The continuum emission shows to be compact 
and seems to be only associated with circumstellar disks.

\item  A molecular LTE  model of a
infalling Keplerian disk with an inner cavity of same size agrees very well with the
line observations. 
This model suggests that the circumbinary ring is infalling with a velocity of V$_{inf}$ $\sim$ 1.5 km s$^{-1}$
toward the binary central compact circumstellar disks;
This model in addition confirms that the protostars in the middle of circumbinary ring are 
of intermediate mass.   

\item The ring appears to be not a stationary object. 
The infall velocity we determine is about a factor of 3 slower than the
free-fall velocity corresponding to the dynamical mass. 
This would correspond to a mass accretion rate of about 10$^{-5}$ M$_\odot$/yr.
The infall velocity of 1.5 km s$^{-1}$ is reached at a distance of the reference radius, 720 AU.

\item We suggest that dust emission associated with Ori 139-409 seems to be mostly 
arising from the circumstellar disks with not strong contribution from the molecular gas ring;

\item A simple comparison with other classical molecular
dusty rings (e.g. GG Tau, UZ Tau, and UY Aur) suggests that {\it Ori 139-409} could be one of the youngest 
circumbinary ring found up to date. {\it Ori 139-409} is related with a class 0 object;

\item The millimeter continuum emission from those rings in Taurus and Auriga 
complexes are very strong compared to what is observed in the {\it Ori 139-409} system, 
while the molecular emission seems to go in the
other sense, with the molecular emission from {\it Ori 139-409} stronger;

\end{itemize}

All these topics are amenable to undertake more high angular observations that aim
at the problem of multiple star formation, such as the presence of accretion streamers,
a connection between the circumbinary and circumstellar disks.

Facilities: {\it SMA}

\bibliographystyle{mn2e}
\bibliography{biblio}

\begin{thebibliography}{}

\bibitem[\protect\citeauthoryear{{Andrews} \& {Williams}}{{Andrews} \&
  {Williams}}{2005}]{And2005}
{Andrews} S.~M.,  {Williams} J.~P.,  2005, \apjl, 619, L175

\bibitem[\protect\citeauthoryear{{Dutrey}}{{Dutrey}}{2001}]{dutrey2001}
{Dutrey} A.,  2001, in {Zinnecker} H.,  {Mathieu} R.,  eds, The Formation of
  Binary Stars Vol.~200 of IAU Symposium, {Interferometric Observations of
  Disks around PMS Binary Stars}.
pp 219--+

\bibitem[\protect\citeauthoryear{{Duvert}, {Dutrey}, {Guilloteau}, {Menard},
  {Schuster}, {Prato} \& {Simon}}{{Duvert} et~al.}{1998}]{Duv1998}
{Duvert} G.,  {Dutrey} A.,  {Guilloteau} S.,  {Menard} F.,  {Schuster} K.,
  {Prato} L.,    {Simon} M.,  1998, \aap, 332, 867

\bibitem[\protect\citeauthoryear{{Eisner} \& {Carpenter}}{{Eisner} \&
  {Carpenter}}{2006}]{eis2006}
{Eisner} J.~A.,  {Carpenter} J.~M.,  2006, \apj, 641, 1162

\bibitem[\protect\citeauthoryear{{Eisner}, {Plambeck}, {Carpenter}, {Corder},
  {Qi} \& {Wilner}}{{Eisner} et~al.}{2008}]{eis2008}
{Eisner} J.~A.,  {Plambeck} R.~L.,  {Carpenter} J.~M.,  {Corder} S.~A.,  {Qi}
  C.,    {Wilner} D.,  2008, \apj, 683, 304

\bibitem[\protect\citeauthoryear{{Gaume}, {Wilson}, {Vrba}, {Johnston} \&
  {Schmid-Burgk}}{{Gaume} et~al.}{1998}]{Gau1998}
{Gaume} R.~A.,  {Wilson} T.~L.,  {Vrba} F.~J.,  {Johnston} K.~J.,
  {Schmid-Burgk} J.,  1998, \apj, 493, 940

\bibitem[\protect\citeauthoryear{{Ghez}, {Emerson}, {Graham}, {Meixner} \&
  {Skinner}}{{Ghez} et~al.}{1994}]{Ghez1994}
{Ghez} A.~M.,  {Emerson} J.~P.,  {Graham} J.~R.,  {Meixner} M.,    {Skinner}
  C.~J.,  1994, \apj, 434, 707

\bibitem[\protect\citeauthoryear{{Guilloteau}, {Dutrey} \&
  {Simon}}{{Guilloteau} et~al.}{1999}]{Gui1999}
{Guilloteau} S.,  {Dutrey} A.,    {Simon} M.,  1999, \aap, 348, 570

\bibitem[\protect\citeauthoryear{{Hioki}, {Itoh}, {Oasa}, {Fukagawa}, {Kudo},
  {Mayama}, {Funayama}, {Hayashi}, {Hayashi}, {Pyo}, {Ishii}, {Nishikawa} \&
  {Tamura}}{{Hioki} et~al.}{2007}]{Hio2007}
{Hioki} T.,  {Itoh} Y.,  {Oasa} Y.,  {Fukagawa} M.,  {Kudo} T.,  {Mayama} S.,
  {Funayama} H.,  {Hayashi} M.,  {Hayashi} S.~S.,  {Pyo} T.-S.,  {Ishii} M.,
  {Nishikawa} T.,    {Tamura} M.,  2007, \aj, 134, 880

\bibitem[\protect\citeauthoryear{{Ho}, {Moran} \& {Lo}}{{Ho}
  et~al.}{2004}]{Hoetal2004}
{Ho} P.~T.~P.,  {Moran} J.~M.,    {Lo} K.~Y.,  2004, \apjl, 616, L1

\bibitem[\protect\citeauthoryear{{Launhardt}}{{Launhardt}}{2004}]{Lau2004}
{Launhardt} R.,  2004, in {Burton} M.~G.,  {Jayawardhana} R.,   {Bourke} T.~L.,
   eds, Star Formation at High Angular Resolution Vol.~221 of IAU Symposium,
  {Observations of Binary Protostellar Systems}.
pp 213--+

\bibitem[\protect\citeauthoryear{{Menten}, {Reid}, {Forbrich} \&
  {Brunthaler}}{{Menten} et~al.}{2007}]{Mentenetal2007}
{Menten} K.~M.,  {Reid} M.~J.,  {Forbrich} J.,    {Brunthaler} A.,  2007, \aap,
  474, 515

\bibitem[\protect\citeauthoryear{{Monin}, {Clarke}, {Prato} \&
  {McCabe}}{{Monin} et~al.}{2007}]{mo2007}
{Monin} J.-L.,  {Clarke} C.~J.,  {Prato} L.,    {McCabe} C.,  2007, in
  {Reipurth} B.,  {Jewitt} D.,   {Keil} K.,  eds, Protostars and Planets V
  {Disk Evolution in Young Binaries: From Observations to Theory}.
pp 395--409

\bibitem[\protect\citeauthoryear{{Pichardo}, {Sparke} \& {Aguilar}}{{Pichardo}
  et~al.}{2005}]{Picha2005}
{Pichardo} B.,  {Sparke} L.~S.,    {Aguilar} L.~A.,  2005, \mnras, 359, 521

\bibitem[\protect\citeauthoryear{{Roddier}, {Roddier}, {Northcott}, {Graves} \&
  {Jim}}{{Roddier} et~al.}{1996}]{Ro1996}
{Roddier} C.,  {Roddier} F.,  {Northcott} M.~J.,  {Graves} J.~E.,    {Jim} K.,
  1996, \apj, 463, 326

\bibitem[\protect\citeauthoryear{{Schmid-Burgk}, {Guesten}, {Mauersberger},
  {Schulz} \& {Wilson}}{{Schmid-Burgk} et~al.}{1990}]{Schmid-Burgketal1990}
{Schmid-Burgk} J.,  {Guesten} R.,  {Mauersberger} R.,  {Schulz} A.,    {Wilson}
  T.~L.,  1990, \apjl, 362, L25

\bibitem[\protect\citeauthoryear{{Scoville}, {Carlstrom}, {Chandler},
  {Phillips}, {Scott}, {Tilanus} \& {Wang}}{{Scoville}
  et~al.}{1993}]{Scovilleetal1993}
{Scoville} N.~Z.,  {Carlstrom} J.~E.,  {Chandler} C.~J.,  {Phillips} J.~A.,
  {Scott} S.~L.,  {Tilanus} R.~P.~J.,    {Wang} Z.,  1993, \pasp, 105, 1482

\bibitem[\protect\citeauthoryear{{Simon}, {Dutrey} \& {Guilloteau}}{{Simon}
  et~al.}{2000}]{Simon2000}
{Simon} M.,  {Dutrey} A.,    {Guilloteau} S.,  2000, \apj, 545, 1034

\bibitem[\protect\citeauthoryear{{Smith}, {Bally}, {Shuping}, {Morris} \&
  {Hayward}}{{Smith} et~al.}{2004}]{smi2004}
{Smith} N.,  {Bally} J.,  {Shuping} R.~Y.,  {Morris} M.,    {Hayward} T.~L.,
  2004, \apjl, 610, L117

\bibitem[\protect\citeauthoryear{{Zapata}, {Ho}, {Rodr{\'{\i}}guez}, {Schilke},
  {Kurtz} \& {et al.}}{{Zapata} et~al.}{2007}]{Zapataetal2007}
{Zapata} L.~A.,  {Ho} P.~T.~P.,  {Rodr{\'{\i}}guez} L.~F.,  {Schilke} P.,
  {Kurtz} S.,    {et al.} 2007, \aap, 471, L59

\bibitem[\protect\citeauthoryear{{Zapata}, {Rodr{\'{\i}}guez}, {Ho}, {Zhang},
  {Qi} \& {Kurtz}}{{Zapata} et~al.}{2005}]{Zapataetal2005}
{Zapata} L.~A.,  {Rodr{\'{\i}}guez} L.~F.,  {Ho} P.~T.~P.,  {Zhang} Q.,  {Qi}
  C.,    {Kurtz} S.~E.,  2005, \apjl, 630, L85

\bibitem[\protect\citeauthoryear{{Zapata}, {Rodr{\'{\i}}guez}, {Kurtz} \&
  {O'Dell}}{{Zapata} et~al.}{2004}]{Zapataetal2004a}
{Zapata} L.~A.,  {Rodr{\'{\i}}guez} L.~F.,  {Kurtz} S.~E.,    {O'Dell} C.~R.,
  2004, \aj, 127, 2252

\bibitem[\protect\citeauthoryear{{Zapata}, {Rodr{\'{\i}}guez}, {Kurtz},
  {O'Dell} \& {Ho}}{{Zapata} et~al.}{2004}]{Zap2004b}
{Zapata} L.~A.,  {Rodr{\'{\i}}guez} L.~F.,  {Kurtz} S.~E.,  {O'Dell} C.~R.,
  {Ho} P.~T.~P.,  2004, \apjl, 610, L121

\bibitem[\protect\citeauthoryear{{Zapata}, {Schmid-Burgk}, {Muders}, {Schilke},
  {Menten} \& {Guesten}}{{Zapata} et~al.}{2009}]{zapa2009}
{Zapata} L.~A.,  {Schmid-Burgk} J.,  {Muders} D.,  {Schilke} P.,  {Menten} K.,
    {Guesten} R.,  2009, ArXiv e-prints

\end{thebibliography}

\end{document}